\documentclass[pdflatex,sn-mathphys-num]{sn-jnl}

\usepackage{graphicx}%
\usepackage{multirow}%
\usepackage{amsmath,amssymb,amsfonts}%
\usepackage{amsthm}%
\usepackage{mathrsfs}%
\usepackage[title]{appendix}%
\usepackage{xcolor}%
\usepackage{textcomp}%
\usepackage{manyfoot}%
\usepackage{booktabs}%
\usepackage{algorithm}%
\usepackage{algorithmicx}%
\usepackage{algpseudocode}%
\usepackage{amsmath}
\usepackage{listings}%
\usepackage{longtable}

\theoremstyle{thmstyleone}%
\theoremstyle{thmstyletwo}%

\theoremstyle{thmstylethree}%

\begin{document}

\title[Article Title]{Algorithms vs. Peers: Shaping Engagement with Novel Content}


\author*[1,2,3]{\fnm{Shan} \sur{Huang}}\email{shanhh@hku.hk}

\author[1]{\fnm{Yi} \sur{Ji}}\email{yi-ji-mkt@connect.hku.hk}

\author[2]{\fnm{Leyu} \sur{Lin}}\email{goshawklin@tencent.com}


\affil*[1]{\orgdiv{Faculty of Business and Economics}, \orgname{The University of Hong Kong}}

\affil[2]{\orgdiv{WeChat Business Group}, \orgname{Tencent}}

\affil[3]{\orgdiv{The Digital Economy Lab, HAI}, \orgname{Stanford University}}



\abstract{
The pervasive rise of digital platforms has reshaped how individuals engage with information, with algorithms and peer influence playing pivotal roles in these processes. This study investigates the effects of algorithmic curation and peer influence through social cues (e.g., peer endorsements) on engagement with novel content. Through a randomized field experiment on WeChat involving over 2.1 million users, we find that while peer-sharing exposes users to more novel content, algorithmic curation elicits significantly higher engagement with novel content than peer-sharing, even when social cues are present. Despite users’ preference for redundant and less diverse content, both mechanisms mitigate this bias, with algorithms demonstrating a stronger positive effect than peer influence. These findings, though heterogeneous, are robust across demographic variations such as sex, age, and network size. Our results challenge concerns about ``filter bubbles" and underscore the constructive role of algorithms in promoting engagement with non-redundant, diverse content.}




\maketitle

\newpage
\section{Introduction}\label{sec1}
Algorithms and peer influence are two primary mechanisms driving online content engagement \cite{Deutsch1955, Gold1956, Bond2017, Baker2016}. Social media platforms such as Facebook and WeChat utilize social cues, such as peer endorsements accompanying peer-shared content, to boost user activity \cite{aral2011, huang2020social}. In contrast, feed algorithms—dominant on newer platforms like TikTok—curate content based on users' behaviors and interests, increasingly shaping contemporary user experiences \cite{garcia2023effect}.

While algorithmic curation excels at personalizing information, it risks isolating users from novel content. This phenomenon fosters ``filter bubbles," where algorithms predominantly present content aligned with users' past behaviors and preferences \cite{Pariser2011, ricci2015recommender}. Similarly, peer-shared content often reflects network homophily—the tendency to associate with like-minded peers \cite{Burt1987}—thereby reinforcing exposure to familiar information \cite{cinelli2021echo}.
However, engagement with novel content is critical for stimulating creativity and innovation \cite{uzzi2013atypical}, reducing information inequality, bridging societal divides, and fostering informed civic participation \cite{Yang2020}. Recognizing these risks, regulatory bodies emphasize promoting content diversity to counteract repetitive informational loops \cite{Pariser2011, Cinelli2021}.

Despite these concerns, both algorithmic curation and peer influence, as typically utilized in peer-shared content, hold significant potential to enhance engagement with novel content, albeit through distinct mechanisms. Algorithms balance exploiting users’ established preferences with exploring new interests. By leveraging machine learning and behavioral insights, algorithms align content with users’ tastes while subtly expanding their informational horizons. Peer influence, in contrast, operates through social cues such as peer endorsements (e.g., peers’ “likes”), which encourage engagement via observational learning and social conformity \cite{Jacoby1972, Friedman1979, Moscovici1985}. Factors such as trust, social identification, and compliance further motivate users to interact with novel content endorsed by their peers \cite{Bakshy2009, Aral2012, Bakshy2012}. Nonetheless, concerns persist that algorithms may intensify “filter bubbles,” restricting exposure to diverse perspectives \cite{Pariser2011}.

Existing research has largely examined algorithms and peer influence separately \cite{Nguyen2014, Han2022, Bakshy2009, huang2020social}, often focusing on feed-ranking algorithms for peer-shared content, particularly on platforms like Facebook \cite{Bakshy2015, Guess2023, Nyhan2023}. This leaves a gap in understanding the role of purely algorithmically curated feeds in shaping user engagement with online content. Studies by Nyhan et al. \cite{Nyhan2023} and Bakshy et al. \cite{Bakshy2015} suggest that while ranking algorithms may reduce exposure to diverse perspectives, they do not necessarily diminish engagement with such content. Despite widespread concerns about ``filter bubbles," the exact role of algorithms in driving this phenomenon remains unclear. Our study addresses this gap, offering large-scale experimental evidence on how algorithms, compared to peer influence, affect engagement with novel content.

Empirically, contexts enabling the evaluation of algorithms and peer influence within the same user group are rare. Observational studies often fail to fully account for platform- and user-specific factors, limiting their ability to isolate causal effects. WeChat, a leading social media platform, provides a unique opportunity to address this challenge. We conducted a large-scale randomized field experiment involving over 2.1 million users on WeChat.
Using peer-shared content without social cues as the baseline, our design isolates the effects of algorithmic curation from peer sharing and examines the impact of displaying social cues (peer influence). Novelty is defined as the non-redundancy and diversity of content relative to users’ historical engagement patterns \cite{Vosoughi2018,aral2023exactly}.

Our findings reveal a nuanced relationship between algorithmic curation, peer influence, and engagement with novel content. Algorithmically curated feeds reduce the novelty of content users are exposed to but significantly increase the novelty of content users engage with, compared to peer-shared content—even when social cues are present. While users tend to prefer redundant and less diverse content for engagement, both algorithms and social cues promote interactions with novel content, with algorithms exerting a significantly stronger positive effect.
These patterns, though heterogeneous, are consistent across demographics, including sex, age, and network size. Collectively, our results challenge the prevailing notion that algorithms inherently constrain content diversity through ``filter bubbles." Instead, they underscore the constructive role of algorithmic curation in fostering engagement with diverse, non-redundant content.

\section{Method}
We collaborated with WeChat to design and conduct a large-scale randomized field experiment on its platform. Specifically, our experiment occurred on WeChat’s “Top Stories” channel, which recommends online articles using two primary mechanisms: algorithmic curation and peer sharing. WeChat’s objective for this channel is to explore algorithmic curation—a newer generation of content delivery strategy—compared to peer-sharing, their initial content delivery approach. This collaboration provides a unique research opportunity to investigate these major mechanisms of content curation. Here, “content” refers specifically to articles distributed within the WeChat ecosystem. A detailed explanation of how WeChat Top Stories leverages both algorithmic curation and peer sharing to disseminate content is provided in Appendix~\ref{app.recommendation_strategy}.

\subsection{Experiment Design}\label{sec.experimentdesign}
A random sample of users were assigned to one of three groups and received the content by three different mechanisms: Group I (algorithm-curated content), Group II (peer-shared content with social cues displayed), and Group III (peer-shared content with social cues hidden, serving as the baseline group). The comparison between Group I and Group III isolates the impact of algorithmic curation in content delivery, the comparison between Group II and Group III identifies the peer influence of displaying social cues with the content, and the comparison between Group I and Group II captures the combined effects of different content curation mechanisms and peer influence.
\footnote{
Notably, the quantity of peer-shared content is constrained by the finite size of a user’s social network, whereas algorithmic curation is virtually unlimited. To maintain service continuity and ensure a consistent volume of recommendations, the platform defaults to algorithmically curated content once all peer-shared articles have been consumed. Consequently, comparisons between Group I (algorithmic curation) and Group II, as well as between Group I and Group III (peer-shared content with and without social cues), provide lower-bound estimates of the between-group differences. The statistical significance of these lower bounds indicates that the actual effects are also significant.
}
\footnote{To accurately measure the impact of social cues with greater statistical power, comparisons between Group II and Group III are restricted to peer-shared content only. This approach provides an estimate of the causal effects of social cues among users whose contacts shared content, thereby giving them the opportunity to be influenced by peers. For users with no peer-shared content, peer influence is effectively zero, allowing this estimate to be regarded as the upper bound of the impact of social cues. If the lower-bound estimate of the impact of algorithmic curation exceeds the upper-bound estimate of the impact of social cues, this relationship is likely to hold for the actual effects.
Furthermore, we conducted robustness checks to validate our findings, focusing on the subset of users who were (potentially) exposed to peer-shared content across all three groups. This allows the exact same user groups when comparing effects of algorithms and peers. The results remain consistent across all analyses. Please find the detailed analyses and results in Appendix \ref{sec.robustness}.}
 
The experiment was conducted over a two-week period, involving 2,141,267 distinct users. Of these, 713,774 users were assigned to Group I, 713,761 to Group II, and 713,732 to Group III. To ensure the validity of the randomized assignments, we conducted a series of randomization checks, including sample ratio mismatch (SRM) analysis, comparisons of mean pre-treatment covariates, and A/A tests. These analyses confirmed the validity of the randomization process (see detailed results in Appendix~\ref{app.randomization_check}).
\footnote{We acknowledge the increased risk of Type I errors when multiple comparisons are made. Adjusted p-values, calculated using methods such as the Bonferroni Correction or the Holm-Bonferroni Method, are typically larger than the original p-values. Therefore, the conclusions drawn from these randomization checks should remain consistent, whether or not adjusted p-values are used.}
Although the experiment was conducted on a social networking platform, the risk of violating the Stable Unit Treatment Value Assumption (SUTVA) \citep{Rubin1990} was minimal. Further details on the SUTVA assessment are provided in Appendix~\ref{app.sutva}.

\begin{figure}[htpb]
\centering
\vspace{1em}
\centerline{Group I \hspace{7em} Group II \hspace{7em} Group III}
\includegraphics[width=0.9\textwidth]{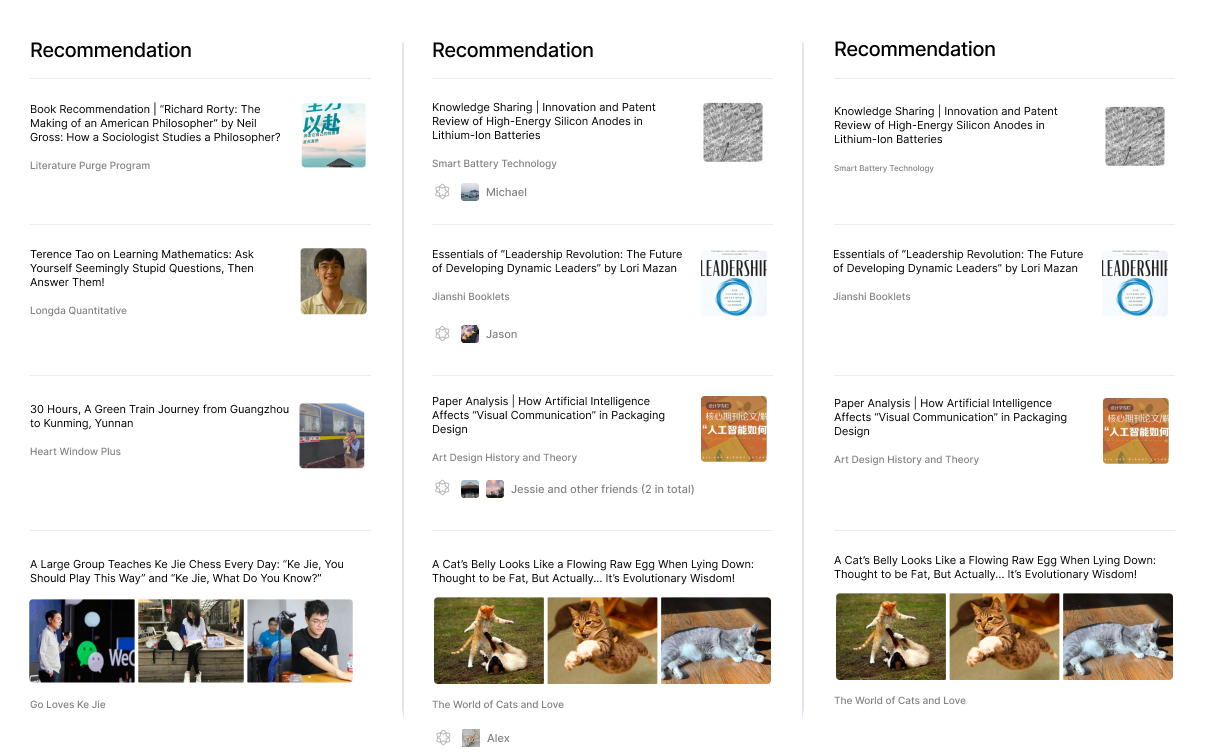}
\caption{\textbf{Experimental design.} Users are randomly assigned to three groups: Group I (exposure to algorithmic curated content), Group II (exposure to peer-shared content with social cues), and Group III (exposure to peer-shared content without social cues).}
\label{fig.design}
\end{figure}

\subsection{Measuring Content Novelty}
Novelty is assessed at the \textit{user-article} level, recognizing that the same content can exhibit varying degrees of novelty for different users. This study examines two dimensions of novelty: \textit{non-redundancy} and \textit{diversity}. Non-redundancy reflects the distinctiveness of content, while diversity captures its contribution to variety. Both dimensions are measured relative to users' historical engagement, specifically their clickthrough activity on WeChat during the month preceding the experiment (pre-experiment period).
Content classification relies on 60 predefined tags (e.g., education, religion, society) developed by the WeChat team for daily content management. Each article is assigned a primary tag based on its predominant theme. A user's historical interests are represented by the tags associated with articles they engaged with during the pre-experiment period. On average, users clicked on 12.5 articles during pre-experiment period, with no statistically significant differences across groups ($p > 0.1$).

\textbf{Content non-redundancy} for an article is defined as 1 if there is no overlap between the article’s tag and the set of tags representing the user’s historical interests. Conversely, if any overlap exists, the content non-redundancy is 0. In robustness checks, we employ alternative continuous measures of non-redundancy, which yield consistent results (see Section~\ref{app.nr_reciprocal} in the Appendix for details).


\textbf{Content diversity} is quantified using Shannon entropy, an information theory metric that measures uncertainty or variability within a dataset \cite{Holtz2020, aral2023exactly, vanherpen2002variety}. Applied to content, Shannon entropy evaluates how evenly content is distributed across categories (tags). For instance, a set of articles evenly spanning multiple topics has higher entropy than one concentrated in a single topic.

We define two types of diversity: baseline diversity and potential diversity. Baseline diversity refers to the Shannon entropy of the content tags associated with a user’s engagements during the month preceding the experiment. Potential diversity represents the Shannon entropy of the information set a user would hold after engaging with a new article. The difference between potential diversity and baseline diversity is termed \textit{marginal diversity}, which quantifies the additional variety introduced by a new article. Marginal diversity thus serves as a measure of how much new content enhances a user’s content diversity. Higher marginal diversity indicates that new content contributes greater variety or novelty. For further details, refer to Section~\ref{app.diversity} in the Appendix.

\section{Results}
\subsection{Novelty in Content Exposure}
We begin our analysis by comparing the average content novelty of articles, to which users are \textit{exposed} under algorithmic curation and peer-sharing, as exposure precedes engagement. Here, content exposure refers to articles delivered to users' WeChat Top Stories homepages and appear on users' smartphone screen, while engagement denotes instances where users click on the articles. 

\begin{figure}[htpb]
\centering
\vspace{0.5em}
\includegraphics[width=0.7\textwidth]{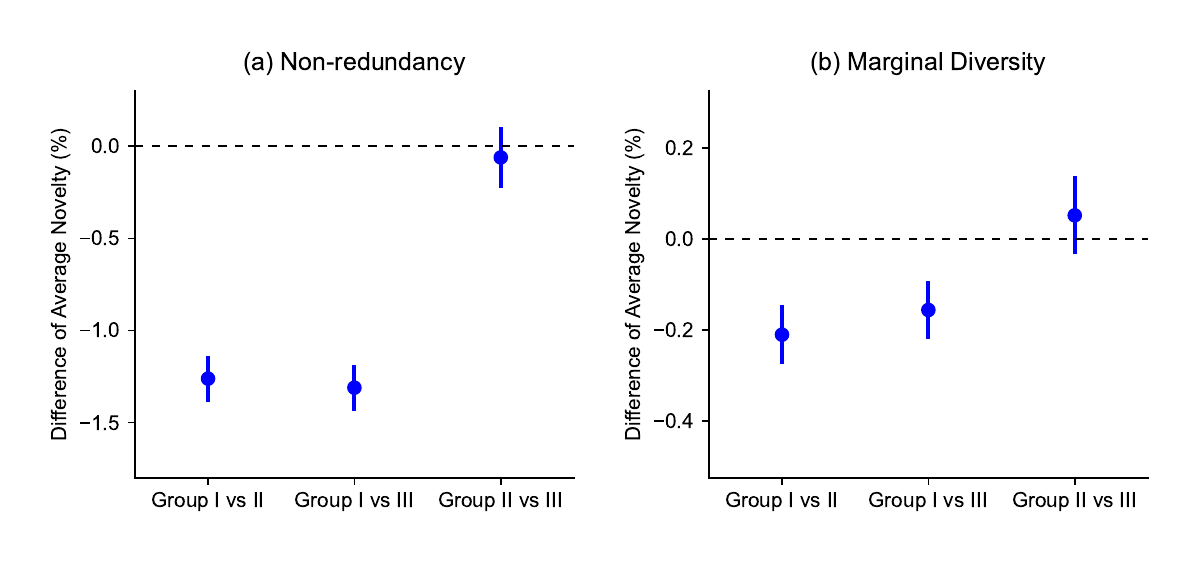}
\caption{\textbf{Average novelty of articles exposed to across groups.} This figure illustrates the comparison of the average novelty level of articles that users are exposed to during the experimental period, across three groups: Group I, with algorithmic curation; Group II, with peer-shared content along with social cues; and Group III (baseline), presenting peer-shared content without social cues. Note that when comparing Group I with Groups II and III, our findings represent the \textit{lower bound} of the effects.}
\label{fig.user-level-exposure}
\end{figure}

Our results show that content shared by peers exhibits significantly higher novelty than algorithmically curated content in terms of content exposure\footnote{All statistical tests in this paper meet the necessary assumptions, including those related to the Central Limit Theorem.}. As illustrated in Figure~\ref{fig.user-level-exposure}, articles exposed through peer-sharing (Groups II and III) display greater average content novelty in both non-redundancy and diversity compared to algorithmic curation (Group I) \((p < 0.01)\). Specifically, the likelihood of encountering non-redundant content is 1.26\% (95\% CI: 1.14\%-1.39\%) higher in Group II (peer-shared content with social cues) and 1.31\% (95\% CI: 1.19\%-1.44\%) higher in Group III (peer-shared content without social cues) relative to Group I. Additionally, marginal diversity, measured by Shannon entropy, is on average 0.0021 (95\% CI: 0.0015-0.0027) higher in Group II and 0.0016 (95\% CI: 0.0009-0.0022) higher in Group III compared to Group I.
As expected, there is no statistically significant difference between Group II and Group III \((p > 0.1)\) in non-redundancy or diversity of an article, as both groups receive peer-shared content and should therefore be indifferent in content exposure.


These findings suggest that algorithmic curation reduces the novelty of content exposure compared to peer-sharing. This aligns with the theory that peer-sharing operates at a local network level, exposing users to content influenced by the diverse interests of their peers. In contrast, algorithmic curation is tailored to the individual level, delivering highly personalized, and potentially narrower, content. Consequently, peers are more likely to share content with greater novelty compared to the individualized recommendations provided by algorithms.

\subsection{Novelty in Content Engagement}\label{sec.contentengagement}
\begin{figure}[htpb]
\centering
\vspace{0.5em}
\includegraphics[width=0.7\textwidth]{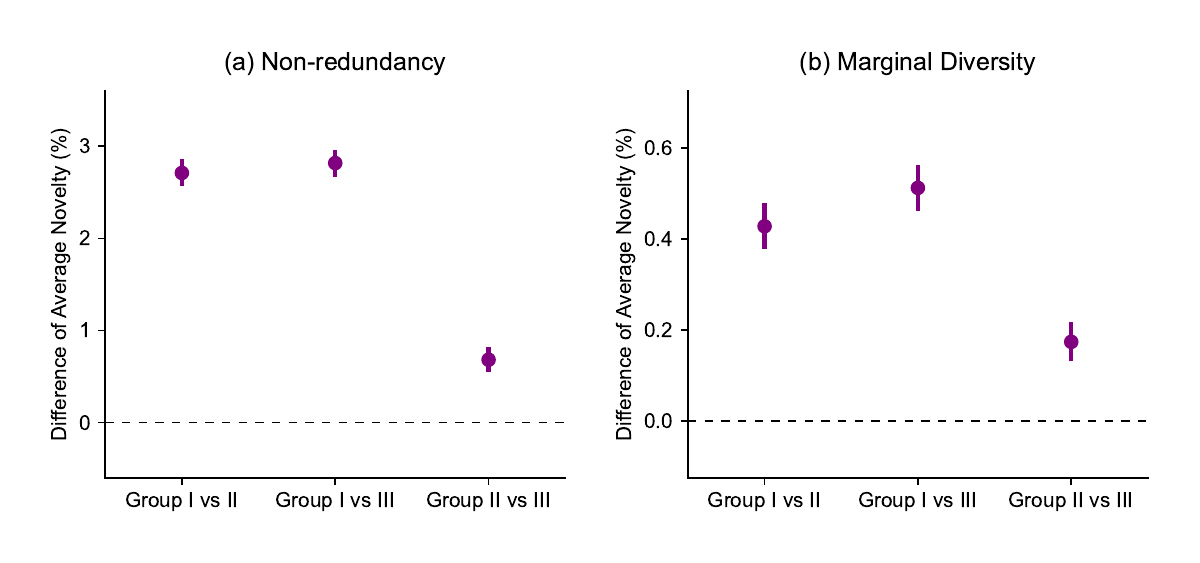}
\caption{\textbf{Average novelty of articles engaged with across groups.} This figure illustrates the comparison of the average novelty level of articles that users engaged with during the experimental period, across three groups: Group I, with algorithmic curation; Group II, with peer-shared content along with social cues; and Group III (baseline), presenting peer-shared content without social cues. 
}
\label{fig.user-level-engagement}
\end{figure}

In contrast to the findings on the novelty of content exposure, our analysis reveals that the average novelty of articles \textit{engaged with} by users is significantly higher under algorithmic curation compared to peer-shared content, even when social cues are displayed (Group II) \((p < 0.01)\). As illustrated in Figure~\ref{fig.user-level-engagement}, this result underscores the positive combined impact of algorithmic content curation and peer influence on engagement with novel content, with algorithmic curation exerting a larger effect than peer influence. In the following sections, we further analyze the distinct impact of algorithmic curation and peer influence to substantiate this finding.


\subsubsection*{The Positive Impact of Algorithmic Curation vs. Peer Influence}
We find that users engage with articles of higher average novelty when content curation is algorithmic (Group I) compared to peer-sharing (Group III). The only distinction between Group I and Group III is the content curation mechanism, as neither includes social cues. Specifically, Figure~\ref{fig.user-level-engagement} shows that the average novelty per article in Group I is 2.82\% (95\% CI: 2.67\%-2.96\%) higher in terms of non-redundancy and 0.0051 (95\% CI: 0.0046-0.0056) higher in terms of diversity compared to Group III \((p < 0.01)\). These findings suggest that while peer-sharing is more likely to expose users to content novel relative to their prior engagement, the novel content recommended by algorithms is more engaging than that shared by peers.

On the other hand, the presence of social cues also significantly increases the average novelty of articles engaged with by users \((p < 0.01)\). Comparing Group II (peer-shared content with social cues) and Group III (peer-shared content without social cues), Figure~\ref{fig.user-level-engagement} illustrates that displaying social cues increases average non-redundancy by 0.68\% (95\% CI: 0.55\%-0.82\%) and diversity by 0.0017 (95\% CI: 0.0013-0.0022).

While both algorithmic curation and social cues significantly enhance the novelty of content that users engage with, algorithmic curation demonstrates a significantly stronger impact than peer influence \((p < 0.01)\). As a result, algorithmically curated content leads to higher novelty in user engagement compared to peer-shared content, even when social cues are present.
Furthermore, the observed divergence between exposed novelty and engaged novelty suggests that users' propensity to engage with novel content, conditional on exposure, differs between algorithmic curation and peer sharing. 

\subsection{Tendency to Engage with Non-Novel Content}\label{sec.contentengagement}
We first examine users' overall intrinsic tendencies to engage with novel content, acknowledging that these behavioral patterns play a critical role in shaping content engagement. We then investigate how these tendencies are influenced by content curation mechanisms and peer influence.
Specifically, we analyze how the non-redundancy and diversity of an article influence users' click-through propensity at the user-article level, conditional on exposure, both on average and across different experimental groups. The analysis employs logistic regression, controlling for user-level variables (e.g., sex, age, city tier, and network degree), a user-article-level variable (the article's ranking for the user), and an article-level variable (the article's total number of clicks, representing its popularity). Detailed analysis is presented in Appendix~\ref{app.reg_results}.

Our results, presented in Table~\ref{tab.general_tendency}, reveal that users consistently prefer engaging with redundant and less diverse content across all experimental groups \((p < 0.01)\). On average, the probability of users clicking on an article significantly decreases as the article’s content becomes non-redundant (Odds Ratio (OR) = 0.518, 95\% CI: 0.515-0.521, \(p < 0.01\)) or more diverse (OR = 0.249, 95\% CI: 0.244-0.254, \(p < 0.01\)).
This preference for less novel content aligns with the theories of \textit{selective exposure} and the \textit{mere exposure effect}. Selective exposure suggests that individuals gravitate toward information that reinforces their existing attitudes and preferences while avoiding content that challenges their views \cite{festinger1962cognitive, nickerson1998confirmation}. Similarly, the mere exposure effect posits that repeated exposure to information fosters familiarity, which enhances preference and encourages engagement with content that users have previously encountered, even superficially \cite{zajonc1968attitudinal}.


\begin{figure}[htpb]
\centering
\vspace{0.5em}
\includegraphics[width=0.6\textwidth]{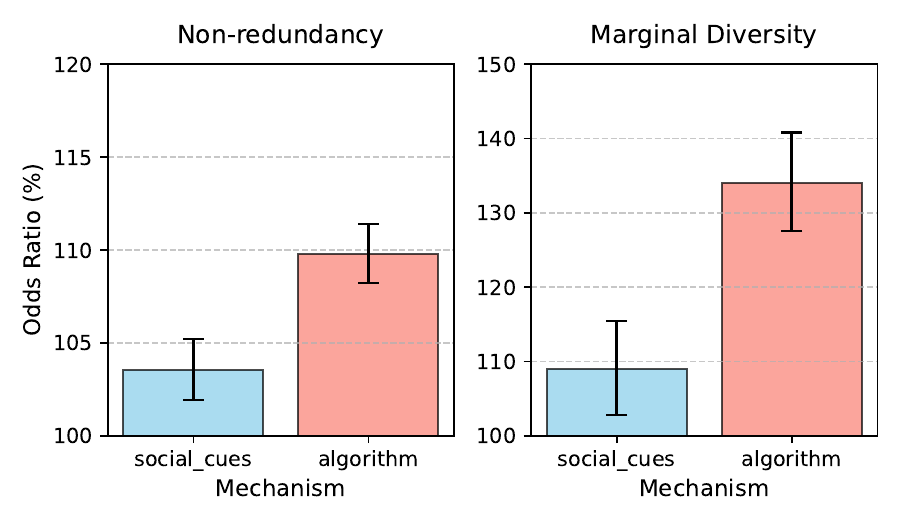}
\caption{\textbf{Effects of social cues and algorithmic curation on users' preference for engaging with non-novel content.} This figure illustrates the odds ratio of engaging with novel content under two mechanisms: social cues and algorithmic curation. The left panel presents the effect in terms of non-redundancy, while the right panel shows the effect in terms of marginal diversity. Algorithmic curation is associated with a significantly higher odds ratio for both measures compared to social cues, with error bars indicating confidence intervals.}
\label{fig.mechanism_comp}
\end{figure}

\subsubsection*{Mitigating Effects of Algorithmic Curation vs. Peer influence}
Although individuals often exhibit a persistent preference for familiar content, our analysis demonstrates that this tendency can be attenuated through algorithmic curation and peer influence. As shown in Figure~\ref{fig.mechanism_comp} and Table~\ref{tab.contentdifference} (Appendix~\ref{app.sec.tendency}), the impact of content non-redundancy (OR = 1.098, 95\% CI: 1.082--1.114, \(p < 0.01\)) and diversity (OR = 1.340, 95\% CI: 1.276--1.408, \(p < 0.01\)) on user engagement propensity is significantly larger and less negative when content is curated algorithmically (Group I) compared to peer sharing (Group III).



Likewise, the effects of content novelty, including both non-redundancy and diversity, exhibit a larger impact when social cues are displayed alongside peer-shared content (Group II vs. Group III; \(p < 0.01\)). Specifically, the inclusion of social cues increases the impact of content non-redundancy (OR = 1.035, 95\% CI: 1.019--1.052, \(p < 0.01\)) and diversity (OR = 1.090, 95\% CI: 1.028--1.155, \(p < 0.01\)) on content engagement. Additionally, algorithmic curation demonstrates a greater impact than social cues in enhancing the influence of content novelty on user engagement. This observation aligns with the finding that content novelty exerts a larger effect on engagement in Group I compared to Group II, where differences exist in both content curation mechanisms and social cues. This is evidenced by both the impact on non-redundancy (OR = 1.090, 95\% CI: 1.074--1.106, \(p < 0.01\)) and diversity (OR = 1.271, 95\% CI: 1.207--1.338, \(p < 0.01\)).

\subsection{Heterogeneous Effects Across Sex, Age and Network Sizes}
\begin{figure}[htpb]
\centering
\vspace{0.5em}
\includegraphics[width=0.7\textwidth]{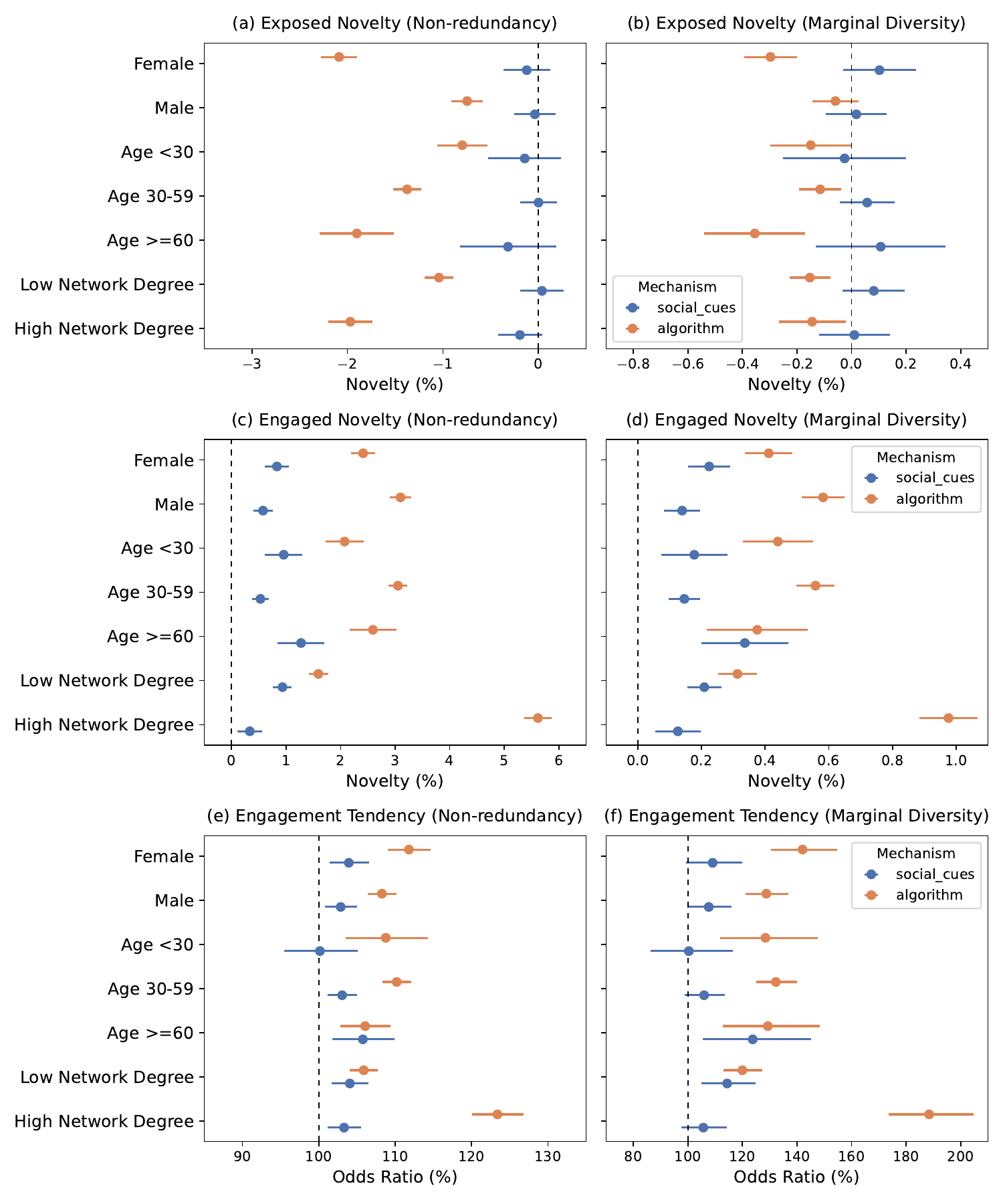}
\caption{\textbf{Heterogeneous effects.} Panels (a) and (b) illustrate the differential impacts of social cues and algorithmic curation on the average novelty levels of articles exposed to distinct user segments. Panels (c) and (d) highlight the effects on the novelty levels of articles actively engaged with by users across these segments. Panels (e) and (f) depict the influence of content novelty on users' propensity to engage with an article under the effects of social cues and algorithmic curation.}
\label{fig.expose-engage-tendency-hetero}
\end{figure}

We expand our analysis to investigate the impact of algorithmic curation and peer influence across demographic groups categorized by sex, age, and network degree—the number of WeChat contacts. A larger network degree, as a typical measure of network centrality, indicates higher social status \cite{burt2000network}. Our results, presented in panels c and d of Figure~\ref{fig.expose-engage-tendency-hetero}, show that all user groups experience positive impacts of algorithmic curation and peer influence on the novelty of content engagement, with algorithmic curation consistently exerting a larger (or at least not smaller) impact than peer influence.
Specifically, algorithmic curation has a stronger effect among males, middle-aged individuals, and those with larger local social networks, on engagement with novel content. In contrast, peer influence (social cues) exerts a greater impact among females, older individuals, and those with smaller local social networks.

These patterns are jointly driven by the heterogeneous effects of algorithmic curation and peer influence on the novelty of content exposure (panels a and b of Figure~\ref{fig.expose-engage-tendency-hetero}) and users’ tendencies to engage with novel content (panels e and f of Figure~\ref{fig.expose-engage-tendency-hetero}).
For example, while algorithmic curation exposes female users to less novel content, it increases their likelihood of engaging with novel content. Similarly, algorithmic curation exposes younger users and those with smaller local social networks to more novel content, but it has a greater impact on novel content engagement among middle-aged users and those with larger local social networks, respectively. Since social cues influence the engaged content novelty only through affecting user behavior rather than content delivery, the results for engaged novelty and the tendency to engage with novel content are generally consistent.

\section{Conclusion}
This study leverages a large-scale randomized field experiment to causally compare algorithmic and peer-driven mechanisms in shaping engagement with novel content. Using peer-shared content as a baseline, we isolate the distinct impacts of algorithmic curation and peer influence mediated by social cues, peer endorsements. While algorithmic curation acts as a content curation mechanism, influencing both the content users are exposed to and their subsequent engagement, peer influence directly shapes users’ perceptions of the same content.

Our findings reveal that while algorithmic curation reduces users’ exposure to novel --- diverse and non-redundant --- content, it significantly outperforms peer influence in driving engagement with novel content. Consequently, algorithmic curation results in higher novelty in engaged content compared to peer-sharing, even when social cues, such as peer endorsements, are present. These findings challenge the widely held assumption that algorithms inherently reinforce “filter bubbles” and instead highlight their constructive role in fostering engagement with novel content.

Algorithmically curated content can be generated at lower costs and in virtually unlimited quantities, whereas peer influence requires substantial investment in establishing and maintaining social networks. Our findings underscore the positive value of algorithms in enhancing content recommendations, offering critical insights for designing recommendation systems that balance personalization with novelty. This work provides essential guidance for platform designers and policymakers in creating more diverse, equitable, and engaging information ecosystems.



\bmhead{Acknowledgements}  
The experiment was conducted following the standard procedures of A/B testing commonly used in tech companies. This study has obtained ethics approval from the University of Hong Kong (HREC's Reference Number: EA250050). Data availability is restricted by a Non-Disclosure Agreement (NDA), the details of which can be provided upon request. All code used for the analysis in the main text and supplementary information is available, and data can be shared in accordance with the terms of the NDA.


\bibliography{sn-bibliography}

\clearpage
\setcounter{page}{1}
\begin{appendices}
\section{Details of Empirical Context}
\label{app.recommendation_strategy}
Peer sharing on WeChat relies on users actively sharing content within their local social networks. Specifically, users express endorsements by tapping the ``wow" button on articles, either under the titles featured in WeChat Top Stories or at the end of the article’s content. Shared items then appear in a user's Top Stories feed \textit{chronologically}, with the most recently shared article displayed at the top. The ranking of peer-shared content is determined solely by the time of sharing, independent of algorithmic processing.
Furthermore, peer-shared articles include social cues indicating which peers have endorsed them. These endorsements are visible under the articles' titles featured in WeChat Top Stories or at the end of the article’s content. This approach provides a unique opportunity to evaluate the effects of purely peer-driven content recommendations, including peer influence, free from algorithmic intervention.

Algorithmic content curation, by contrast, are generated through a two-step process: first, a small subset of items is selected from a larger pool; next, these candidates are ranked based on relevance scores. WeChat Top Stories utilizes a combination of algorithms, including deep neural networks, collaborative filtering, transform learning and etc., to infer user preferences from a vast repository of historical and real-time data. This system delivers personalized recommendations optimized to enhance user engagement metrics. In our experiment, items shared by peers were explicitly excluded from the pool of algorithmically delivered recommendations, ensuring no overlap between the two sets. Consequently, the performance of algorithmic curation in this study might appear slightly diminished compared to their real-world potential. However, this separation was deliberate, allowing for an independent assessment of the effectiveness of algorithmic content curation. 

The algorithm employed in our experiment, similar to those used by leading platforms, is designed to optimize metrics such as clicks, likes, and retention rates, while balancing the trade-off between exploitation and exploration in recommendations. It has been meticulously developed and continuously refined by a dedicated team to ensure its sophistication and effectiveness.

\section{Experiment Randomization Checks}
\label{app.randomization_check}

We conducted a series of randomization checks to validate the group assignments in our experiment. These checks included sample ratio mismatch (SRM), comparisons of mean user characteristics unaffected by experimental treatments, and A/A tests, all of which confirmed the validity of our randomization process.

First, the SRM test results, presented in Table~\ref{tab_srm}, show that the distribution of users among the experimental groups closely matched the expected allocation---groups with equal sample sizes. Specifically, traffic was evenly distributed among each group throughout the experimental period, with no significant differences in the cumulative number of users participating across groups ($p > 0.1$).

Second, we compared the demographics (sex, age, city), network degree (number of WeChat contacts), and engagement metrics with WeChat and WeChat Top Stories (login days, clicks, dwell time, shares) across groups for a four-week period prior to the experiment. These comparisons revealed no significant differences across all measured variables, as shown in Table \ref{tab_randcheck}.

Third, to further ensure the reliability of our experimental platform, we conducted an A/A test. This involved introducing an additional control group, Group IV, created using the same randomization procedure. This group received the same treatment as Group II. The AA test, detailed in Table \ref{tab_aa}, found no statistically significant differences between Groups II and IV in terms of click metrics, dwell time, sharing metrics, or platform retention during the experiment, reinforcing the credibility of our experimental randomization.
\begin{table}[htp]
\centering
\caption{SRM test (cumulative number of users participating in the experiment)}
\begin{tabular}{ccccccc}
\toprule
Date & Group I & Group II & Group III & Group IV &  Chi-squared &  \textit{p}-value \\
\midrule
20220114 & 121503 & 121661 & 121166 & 121838 &     2.013082 & 0.569696 \\
20220115 & 260108 & 260155 & 259760 & 260068 &     0.368594 & 0.946651 \\
20220116 & 333953 & 333858 & 334349 & 333331 &     1.578340 & 0.664311 \\
20220117 & 391258 & 391174 & 391803 & 390788 &     1.341733 & 0.719248 \\
20220118 & 438091 & 438022 & 438303 & 437287 &     1.340160 & 0.719619 \\
20220119 & 478995 & 478757 & 479194 & 477501 &     3.637100 & 0.303411 \\
20220120 & 515881 & 515319 & 516311 & 514319 &     4.313021 & 0.229587 \\
20220121 & 549498 & 549067 & 550040 & 548198 &     3.306796 & 0.346698 \\
20220122 & 579914 & 579595 & 580431 & 578725 &     2.651958 & 0.448453 \\
20220123 & 608582 & 608436 & 609259 & 607341 &     3.112569 & 0.374593 \\
20220124 & 636844 & 637033 & 637293 & 635424 &     3.299844 & 0.347664 \\
20220125 & 663852 & 664046 & 664042 & 661884 &     5.003340 & 0.171553 \\
20220126 & 690435 & 690687 & 690458 & 688336 &     5.272755 & 0.152880 \\
20220127 & 713774 & 713761 & 713732 & 711813 &     3.969591 & 0.264767 \\
\bottomrule
\end{tabular}
\footnotetext{Note: Group IV is designated for the A/A test. Users in Group IV receive the same treatment as those in Group II, where both the items recommended by peers and the names of the peers who ``broadcast" these items are displayed.}
\label{tab_srm}
\end{table}

\begin{table}[htpb]
\centering
\caption{Randomization check using pre-experiment data}
\begin{tabular}{ccccccc}
\toprule 
& \multicolumn{2}{c}{Group I vs Group II} & \multicolumn{2}{c}{Group I vs Group III} & \multicolumn{2}{c}{Group II vs Group III} \\
\cmidrule(r){2-3} \cmidrule(r){4-5} \cmidrule(r){6-7} 
Variable & \textit{t}-statistic & \textit{p}-value & \textit{t}-statistic & \textit{p}-value & \textit{t}-statistic & \textit{p}-value  \\
\midrule
Sex & -0.3825 & 0.7021 & 0.8957 & 0.3704 &  0.5424 &  0.5876 \\
Age & -0.441 &  0.6592 & 0.5463 & 0.5848 &  1.2734 &  0.2029 \\
CityTier &   -0.6597 &  0.5094 & 0.4016 &  0.688 & -1.3831 &  0.1666 \\
NetworkDegree &  0.3128 &  0.7544 & -0.8775 &  0.3802 & -0.1132 &  0.9099 \\
WeChatLoginDays & -0.3614 &  0.7178 & -1.6156 &  0.1062 & -1.7771 &  0.0755\\
TopStoriesLoginDays & -0.5745 &  0.5656 & -0.1896 &  0.8496 &  0.0382 &  0.9695 \\
DummyClick & -1.0983 &  0.2721 & -1.2528 &  0.2103 & -0.363 &  0.7166\\
ClickItemNum & -1.7408 &  0.0817 & -0.4055 &  0.6851 & -0.2013 &  0.8405 \\
ClickRate &  0.3214 &  0.7479 & -0.3476 &  0.7282 &  0.4147 &  0.6784 \\
DwellTime & -0.3594 &  0.7193 & -1.1962 &  0.2316 &  0.0556 &  0.9556\\
ShareItemNum & -0.2445 &  0.8069 & -1.1421 &  0.2534 & -1.7767 &  0.0756 \\
ShareRate & -0.2649 &  0.7911 & -0.9049 &  0.3655 & -1.4895 &  0.1364 \\ 
\bottomrule 
\end{tabular}
\footnotetext{Note: All variables are aggregated over a four-week period preceding the experiment. The mean differences across the three groups for the variables listed in the table are statistically insignificant ($p > 0.05$). These results, particularly when accounting for Type I (false positive) error, provide strong evidence supporting the validity of the randomization in the experiment.}
\label{tab_randcheck} 
\end{table}
\begin{table}[htp]
\centering
\caption{AA test}
\begin{tabular}{ccc}
\toprule
& \multicolumn{2}{c}{Group II vs Group IV} \\
\cmidrule(r){2-3}
Variable & \textit{t-statistics} & \textit{p-value} \\
\midrule
ClickRate   & 0.7358 & 0.4618\\
ClickDays &  0.1253 &  0.9003 \\
ClickItemNum & -0.2841 & 0.7763\\
DwellTime & -0.2597 &  0.7951 \\
LoginDays &  -0.5495 &  0.5827 \\
ShareRate & -0.2001  &  0.8414\\
ShareDay &  -0.0253  & 0.9799 \\
ShareItemNum & -1.7610 &  0.0782 \\
\bottomrule
\end{tabular}
\label{tab_aa}
\end{table}
\section{Stable Unit Treatment Value Assumption}\label{app.sutva}
SUTVA (Stable Unit Treatment Value Assumption) \citep{Rubin1990} is generally supported. There was no treatment spillover among users from different groups; users adhered independently to their experimental conditions. A minor SUTVA concern could be the potential influence of peers' treatments on article-sharing, which might impact articles feeds for Groups II and III. However, the random assignment or exclusion of peers into control or treatment groups ensures experiment's internal validity - the treatments' effects on peers are equally distributed across groups. Further, this impact also should be minimal: less than 10\% of focal users' peers were involved in the experiment, their sharing rate and the click rate on these shared articles by focal users was less than 10\%. 


\section{Alternative Measurement of Non-redundancy}
\label{app.nr_reciprocal}
\begin{figure}[htpb]
\centering
\includegraphics[width=0.7\textwidth]{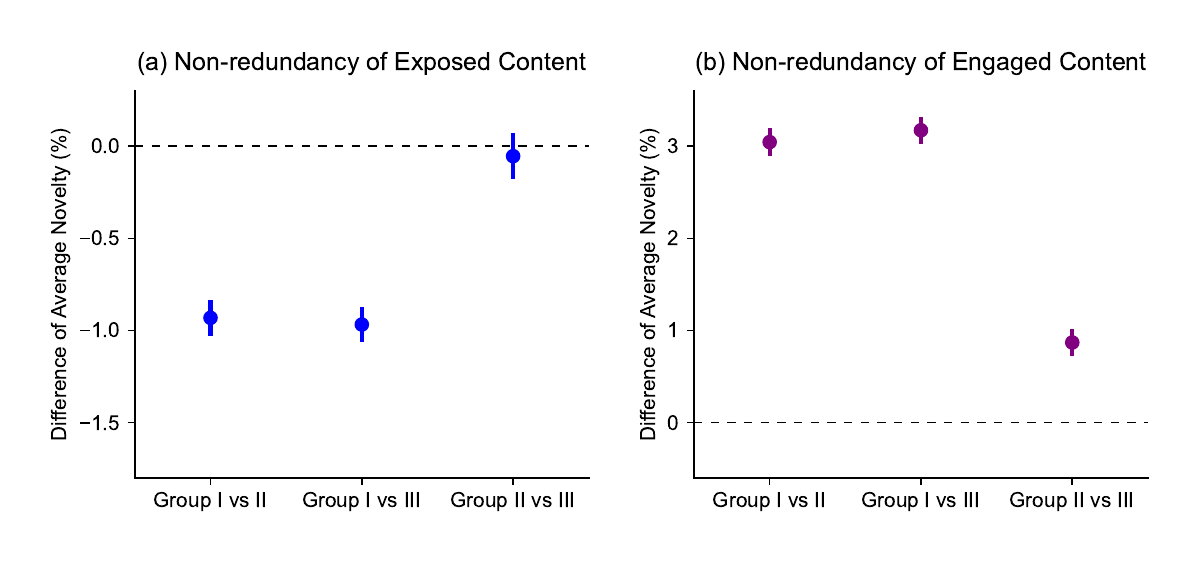}
\caption{\textbf{Average novelty of articles exposed to and engaged with across groups (Reciprocal Method of Non-redundancy).} The figure compares the average novelty levels of articles encountered (panel a) and engaged with (panel b) by users during the experimental period, across three groups: Group I, with algorithmic curation; Group II, with peer-shared content along with social cues; and Group III (baseline), presenting peer-shared content without social cues. Note that when comparing Group I with Groups II and III, our findings represent the \textit{lower bound} of the effects.}
\label{fig.user-level-exposure-engagement-Reciprocal}
\end{figure}

\begin{figure}[htpb]
\centering
\includegraphics[width=0.35\textwidth]{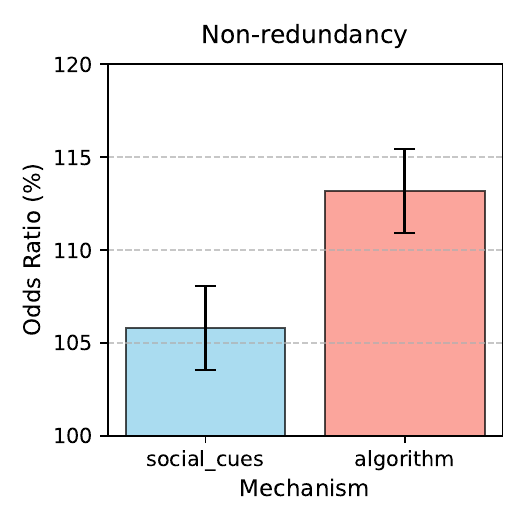}
\caption{\textbf{Effects of social cues and algorithmic curation on users' preference for engaging with non-novel content (Reciprocal Method of Non-redundancy).} This figure illustrates the odds ratio of engaging with novel content under two mechanisms: social cues and algorithmic curation. The left panel presents the effect in terms of non-redundancy, while the right panel shows the effect in terms of marginal diversity. Algorithmic curation is associated with a significantly higher odds ratio for both measures compared to social cues, with error bars indicating confidence intervals.}
\label{fig.mechanism_comp_reciprocal}
\end{figure}

We adopt an alternative measure of content non-redundancy. Instead of using a binary variable, we utilize the reciprocal of the number of articles plus one (1/(n+1)) that a user engaged with under the same tag in the previous month as our measure. This metric, as a continuous measure, captures the diminishing novelty effects and offers a more sensitive assessment compared to the prior definition of non-redundancy as a binary variable. The results presented in Figures~\ref{fig.user-level-exposure-engagement-Reciprocal} and~\ref{fig.mechanism_comp_reciprocal}
are consistent with and further reinforce our main findings.

\newpage
\section{Robustness Check} \label{sec.robustness}
To address the possibility that some users may not receive peer-shared content from their WeChat contacts—likely due to small local networks or less active contacts—we conducted additional analyses among users likely to receive peer-shared content. As this condition applies equally across all experimental groups, the randomization and causal identification of our experiment remain valid.

Specifically, we developed a logistic regression model using pre-treatment variables to estimate whether a user receives peer-shared content. The model incorporates users’ pre-treatment behavioral history (i.e., the number of peer-shared articles received in the month prior to the experiment) and user characteristics (i.e., sex, age, city tier, and network degree). Data from Groups II and III, where the ground truth of receiving peer-shared content is known, were used for model development.
We randomly split the users from these two groups into a training set (75\%) and a testing set (25\%). After training the model on the training set, we achieved a training accuracy of 82.4\% and a testing accuracy of 82.3\%.

We then applied this model to all three user groups to identify a sample of users predicted to receive peer-shared content. Although the ground truth was observed in Groups II and III, we still applied the predictive model across all groups to ensure sample balance across experimental groups and preserve the causal identification of the experiment.

\begin{figure}[htpb]
\centering
\includegraphics[width=0.7\textwidth]{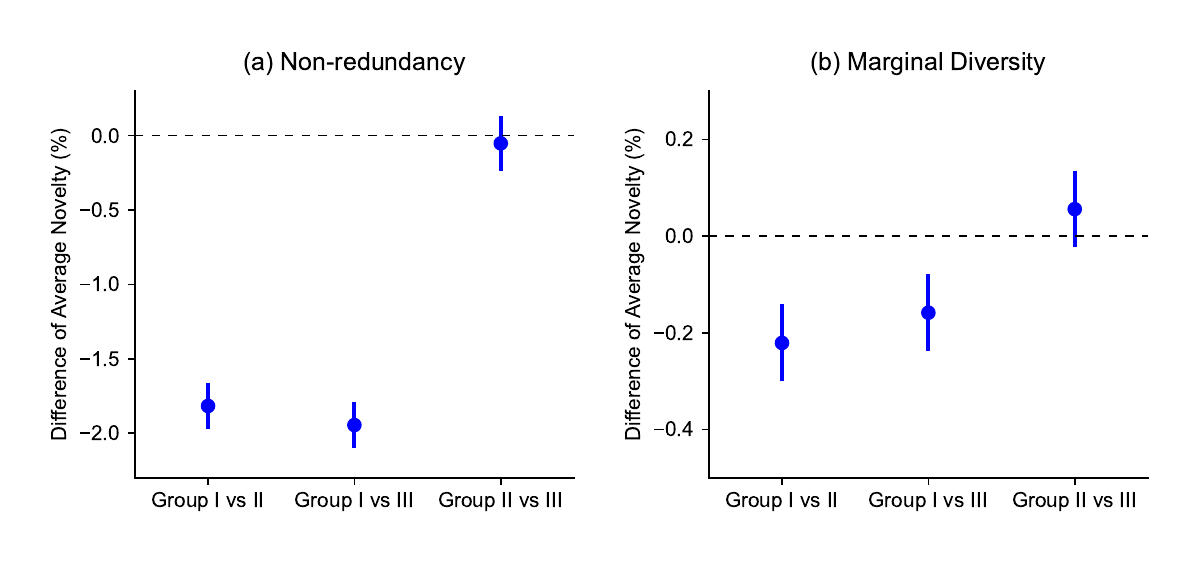}
\caption{\textbf{Average novelty of articles exposed to across groups (Robustness).} This figure illustrates the comparison of the average novelty level of articles that users are exposed to during the experimental period, across three groups: Group I, with algorithmic curation; Group II, with peer-shared content along with social cues; and Group III (baseline), presenting peer-shared content without social cues. Note that when comparing Group I with Groups II and III, our findings represent the \textit{lower bound} of the effects.}
\label{fig.user-level-exposure-match}
\end{figure}

\begin{figure}[htpb]
\centering
\includegraphics[width=0.7\textwidth]{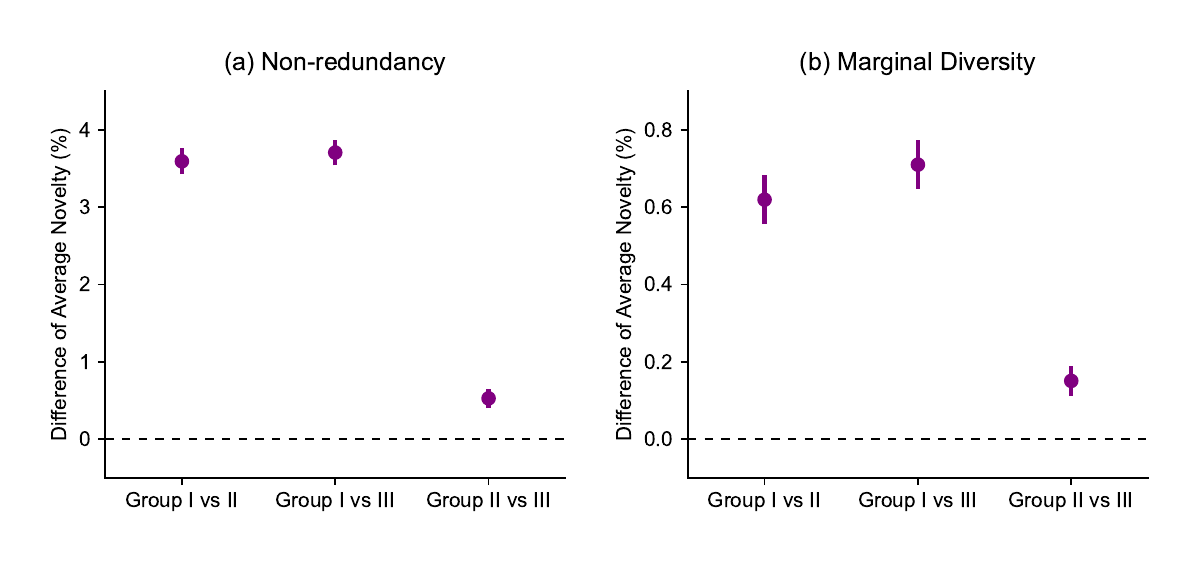}
\caption{\textbf{Average novelty of articles engaged with across groups (Robustness).} This figure illustrates the comparison of the average novelty level of articles that users engaged with during the experimental period, across three groups: Group I, with algorithmic curation; Group II, with peer-shared content along with social cues; and Group III (baseline), presenting peer-shared content without social cues. Note that when comparing Group I with Groups II and III, our findings represent the \textit{lower bound} of the effects.}
\label{fig.user-level-engagement-match}
\end{figure}

Finally, we repeated the analyses on this newly selected sample. The results, presented in Figures~\ref{fig.user-level-exposure-match} and \ref{fig.user-level-engagement-match} are consistent with our main findings.

\begin{figure}[htpb]
\centering
\includegraphics[width=0.6\textwidth]{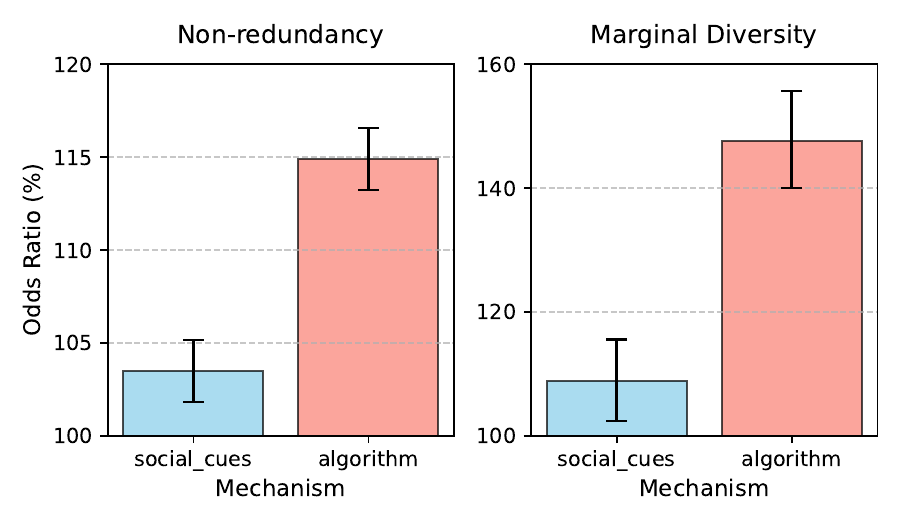}
\caption{\textbf{Effects of social cues and algorithmic curation on users' preference for engaging with non-novel content (Robustness).} This figure illustrates the odds ratio of engaging with novel content under two mechanisms: social cues and algorithmic curation. The left panel presents the effect in terms of non-redundancy, while the right panel shows the effect in terms of marginal diversity. Algorithmic curation is associated with a significantly higher odds ratio for both measures compared to social cues, with error bars indicating confidence intervals.}
\label{fig.mechanism_comp_match}
\end{figure}
\newpage
\section{Content Diversity}\label{app.diversity}
We measure \textit{content diversity} using \textit{Shannon entropy}, a concept from information theory that quantifies the uncertainty or variability within a set of information \cite{Holtz2020, aral2023exactly, vanherpen2002variety}. When applied to content, Shannon entropy captures how evenly the content is distributed across different tags. A set of articles evenly spread across multiple topics has higher entropy than a set dominated by a single topic.

The \textbf{Shannon entropy} \( H \) of a set of content tags \( \{t_i\} \) is defined as:
\begin{equation}
H = -\sum_{i=1}^n p_i \log p_i,
\end{equation}
where \( p_i \) represents the probability (or proportion) of the tag \( t_i \) within the set, and \( n \) denotes the total number of distinct tags. For users who did not engage with any content last month, the Shannon entropy is defined as zero.

\noindent
\textbf{1. Baseline Diversity} (\( H_{\text{baseline}} \)):  
The Shannon entropy of the set of content tags a user engaged with during the month preceding the experiment:
\begin{equation}
H_{\text{baseline}} = -\sum_{i=1}^n p_i^{\text{baseline}} \log p_i^{\text{baseline}}
\end{equation}
\textbf{2. Potential Diversity} (\( H_{\text{potential}} \)):
The Shannon entropy of the information set given a user reads a new article:
\begin{equation}
H_{\text{potential}} = -\sum_{i=1}^n p_i^{\text{potential}} \log p_i^{\text{potential}}
\end{equation}
\textbf{3. Marginal Diversity} (\( \Delta H \)): 
The difference between potential diversity and baseline diversity, representing the additional diversity introduced by the new content:
\begin{equation}
\Delta H = H_{\text{potential}} - H_{\text{baseline}}
\end{equation}



\section{Regression Results}
\label{app.reg_results}


\subsection{Users' Tendency to Engage with More Novel Content Across Groups}
\label{app.sec.tendency}

We explore the general tendency of users to engage with more novel content. Specifically, we examine how the novelty — measured in terms of non-redundancy and diversity — of an article affects users' likelihood of clicking through it when they are exposed to it. To model the binary outcomes of user engagement, we employ a logistic regression model as follows:
\begin{equation}\label{eq1}
       logit(Pr (Y_{ij})) = \alpha + \beta_1 Novelty_{ij} + \beta_2 C_{ij} + \epsilon_{ij},
\end{equation} where $Y_{ij}$ represents the dependent variable, a binary outcome indicating whether $user_i$ clicked on $article_j$. The variable $Novelty_{ij}$ is a binary indicator denoting whether $article_j$ is considered novel for $user_i$. 

\begin{sidewaystable}[htp]
\centering
\caption{The impact of novelty on user content engagement across groups}
\begin{tabular}{c@{\hskip 2.5pt}c@{\hskip 2.5pt}c@{\hskip 2.5pt}c@{\hskip 2.5pt}c@{\hskip 2.5pt}c@{\hskip 2.5pt}c@{\hskip 2.5pt}c@{\hskip 2.5pt}c@{\hskip 2.5pt}c@{\hskip 2.5pt}c@{\hskip 2.5pt}c@{\hskip 2.5pt}c@{\hskip 2.5pt}}
\toprule 
& \multicolumn{12}{c}{Engagement}
\\
\cmidrule(r){2-13}
& \multicolumn{4}{c}{Group I} & \multicolumn{4}{c}{Group II} & \multicolumn{4}{c}{Group III} \\
\cmidrule(r){2-5} \cmidrule(r){6-9} \cmidrule(r){10-13}
& (1) & (2) & (3) & (4) & (5) & (6) & (7) & (8) & (9) & (10) & (11) & (12) \\
\midrule 
NR & $-0.560^{***}$ & $-0.589^{***}$ & & & $-0.620^{***}$ & $-0.681^{***}$ & & & $-0.632^{***}$ & $-0.689^{***}$ & & \\
& (0.006) & (0.005) & & & (0.006) & (0.005) & & & (0.007) & (0.005) & & \\
Diversity & & & $-1.154^{***}$ & $-1.220^{***}$ & & & $-1.334^{***}$ & $-1.443^{***}$ & & & $-1.392^{***}$ & $-1.501^{***}$ \\
& & & (0.021) & (0.017) & & & (0.022) & (0.020) & & & (0.022) & (0.018) \\
Constant & $-1.597^{***}$ & $-2.132^{***}$ & $-1.717^{***}$ & $-2.703^{***}$ & $-1.668^{***}$ & $-2.262^{***}$ & $-1.798^{***}$ & $-2.941^{***}$ & $-1.669^{***}$ & $-2.400^{***}$ & $-1.803^{***}$ & $-3.133^{***}$ \\
& (0.006) & (0.089) & (0.005) & (0.093) & (0.006) & (0.160) & (0.005) & (0.160) & (0.006) & (0.081) & (0.005) & (0.087) \\
Controls & & Yes & & Yes & & Yes & & Yes & & Yes & & Yes \\
\midrule 
Observations & 32,148,613 & 32,148,613 & 32,148,613 & 32,148,613 & 32,778,408 & 32,778,408 & 32,778,408 & 32,778,408 & 32,606,951 & 32,606,951 & 32,606,951 & 32,606,951 \\
Pseudo $R^{2}$ & 0.010 & 0.063 & 0.003 & 0.056 & 0.011 & 0.047 & 0.004 & 0.039 & 0.012 & 0.048 & 0.004 & 0.040 \\
\bottomrule 
\end{tabular}
\footnotetext{Note: Clustered standard errors at the user level in parentheses. Stars indicate the significance level based on $p$ values: ${}^{*} p < 0.1$; ${}^{**} p < 0.05$; ${}^{***} p < 0.01$.}  
\label{tab.general_tendency}
\end{sidewaystable}
\newpage

\subsection{Impact of Peer Influence: Group II vs. III}
We examine whether the display of social cues increases users' propensity to engage with novel content, by comparing Group II (peer-shared content with social cues) and Group III (peer-shared content without social cues). Theoretical frameworks suggest that social influence, driven by mechanisms such as social proof and social conformity, can motivate users to interact with unfamiliar information. For this analysis, we employ a logistic regression model with an interaction term as outlined in Model \ref{eq2}.
\begin{align}\label{eq2}
logit(Pr(Y_{ij})) = \alpha &+ \beta_1 Group_i + \beta_2 Novelty_{ij} + \gamma_1 Group_i\times Novelty_{ij} \nonumber \\ 
& + \beta_3 C_{ij} + \epsilon_{ij},
\end{align}
where $Group_i$ is a dummy indicator for whether $user_i$ is in Group I or Group II. 

The results shown in Table \ref{tab.socialcue} corroborate theoretical predictions, demonstrating that social cues significantly increase users' propensity to engage with more non-redundant or more diverse content and reduce their inclination to interact with non-novel content ($p < 0.001$). However, the impact of the content difference is larger than that of social cues, resulting in a significantly higher tendency among users to engage with novel information under algorithmic delivery (Group I) compared to peer-shared delivery even with social cues displayed (Group II). These results are consistent with and further explained our findings in Section \ref{sec.contentengagement}.

\newpage
\subsection{Impact of Algorithmic Curation: Group I vs. III}
We investigate how content curation by algorithms, compared to peer-sharing, influences users' tendency to engage with more novel content—defined as less redundant or more diverse—by comparing Group I and Group III. In these two groups, articles are delivered through distinct recommendation systems, with neither group receiving social cues. The logistic regression model outlined in Model \ref{eq2} is applied to analyze this relationship.

The results, presented in Table \ref{tab.contentdifference}, show that the impact of non-redundancy and diversity on users' propensity to engage with content is significantly larger and less negative under algorithmic curation (Group I) compared to peer-sharing without social cues (Group III) ($p < 0.001$). These findings highlight the beneficial impact of algorithmic curation over peer-sharing in encouraging users to engage with more novel content.
\begin{table}[htp]
\centering
\caption{Users' tendency to engage with more novel content between Group II vs. III}
\begin{tabular}{ccccc}
\toprule 
& \multicolumn{4}{c}{Engagement (Group II vs Group III)} \\
\cmidrule(r){2-5}
& (1) & (2) & (3) & (4) \\
\midrule 
$\text{SocialCue}$ $\times$ NR & $0.042^{***}$ & $0.035^{***}$ & & \\
& (0.008) & (0.008) & & \\
NR & $-0.919^{***}$ & $-0.888^{***}$ & & \\
& (0.006) & (0.006) & & \\
$\text{SocialCue}$ $\times$ Diversity & & & $0.092^{***}$ & $0.086^{***}$ \\
& & & (0.029) & (0.030) \\
Diversity & & & $-1.710^{***}$ & $-1.679^{***}$ \\
& & & (0.021) & (0.021) \\
$\text{SocialCue}$ & $0.069^{***}$ & $0.074^{***}$ & $0.084^{***}$ & $0.086^{***}$ \\
& (0.007) & (0.006) & (0.006) & (0.006) \\
Constant & $-1.993^{***}$ & $-3.037^{***}$ & $-2.240^{***}$ & $-3.435^{***}$ \\
& (0.005) & (0.031) & (0.004) & (0.032) \\
Controls & & Yes & & Yes \\
\midrule 
Observations & 7,529,256 & 7,529,256 & 7,529,256 & 7,529,256 \\
Pseudo $R^{2}$ & 0.025 & 0.044 & 0.008 & 0.029 \\
\bottomrule 
\end{tabular}
\footnotetext{Note: Clustered standard errors at the user level in parentheses. Stars indicate the significance level based on $p$ values: ${}^{*} p < 0.1$; ${}^{**} p < 0.05$; ${}^{***} p < 0.01$.}

\label{tab.socialcue}
\end{table}

\begin{table}[htp]
\centering
\caption{Users' tendency to engage with more novel content between Group I vs. III}
\begin{tabular}{ccccc}
\toprule 
 & \multicolumn{4}{c}{Engagement (Group I vs Group III)}
 \\
 \cmidrule(r){2-5} 
 & (1) & (2) & (3) & (4) \\
\midrule 
$\text{Algorithm}$ $\times$ NR & $0.073^{***}$ & $0.093^{***}$ & & \\
& (0.009) & (0.007) & & \\
NR & $-0.632^{***}$ & $-0.689^{***}$ & & \\
& (0.007) & (0.005) & & \\
$\text{Algorithm}$ $\times$ Diversity & & & $0.237^{***}$ & $0.293^{***}$ \\
& & & (0.030) & (0.025) \\
Diversity & & & $-1.391^{***}$ & $-1.513^{***}$ \\
& & & (0.022) & (0.018) \\
$\text{Algorithm}$ & $0.072^{***}$ & $0.077^{***}$ & $0.086^{***}$ & $0.098^{***}$ \\
& (0.008) & (0.007) & (0.007) & (0.006) \\
Constant & $-1.669^{***}$ & $-2.296^{***}$ & $-1.803^{***}$ & $-2.671^{***}$ \\
& (0.006) & (0.030) & (0.005) & (0.030) \\
Controls & & Yes & & Yes \\
\midrule 
Observations & 64,755,564 & 64,755,564 & 64,755,564 & 64,755,564 \\
Pseudo $R^{2}$ & 0.011 & 0.055 & 0.004 & 0.047 \\
\bottomrule 
\end{tabular}
\footnotetext{Note: Clustered standard errors at the user level in parentheses. Stars indicate the significance level based on $p$ values: ${}^{*} p < 0.1$; ${}^{**} p < 0.05$; ${}^{***} p < 0.01$.}    
\label{tab.contentdifference}
\end{table}

\begin{sidewaystable}
\centering
\caption{Heterogeneous effects of social cues on novel information engagement}
\begin{tabular}{cccccccc}
\toprule 
& \multicolumn{7}{c}{Engagement (Group II vs Group III)} \\
\cmidrule(r){2-7} 
& \multicolumn{2}{c}{(1) Sex} & \multicolumn{3}{c}{(2) Age}  & \multicolumn{2}{c}{(3) Network Degree} \\
\cmidrule(r){2-3} \cmidrule(r){4-6} \cmidrule(r){7-8}
& Female & Male & $<30$ & $30-59$ & $\geq60$ & Low & High \\
\midrule 
$\text{SocialCue}$ $\times$ NR & $0.039^{***}$ & $0.028^{***}$ & $0.001$ & $0.030^{***}$ & $0.056^{***}$ & $0.040^{***}$ & $0.033^{***}$ \\
& (0.013) & (0.010) & (0.024) & (0.009) & (0.020) & (0.012) & (0.011) \\
\midrule 
$\text{SocialCue}$ $\times$ Diversity & $0.087^{*}$ & $0.074^{*}$ & $0.003$ & $0.058$ & $0.213^{***}$ & $0.134^{***}$ & $0.055$ \\
& (0.048) & (0.038) & (0.077) & (0.035) & (0.082) & (0.044) & (0.040) \\
\midrule 
Observations & 2,822,573 & 4,706,683 & 917,954 & 5,645,800 & 965,502 & 3,140,512 & 4,388,744 \\
\bottomrule 
\end{tabular}
\footnotetext{Note: Clustered standard errors at the user level in parentheses. Stars indicate the significance level based on $p$ values: ${}^{*} p < 0.1$; ${}^{**} p < 0.05$; ${}^{***} p < 0.01$.}  
\label{tab.hetero_socialcue}
\end{sidewaystable}

\begin{sidewaystable}
\centering
\caption{Heterogeneous effects of content difference on novel information engagement}
\begin{tabular}{cccccccc}
\toprule 
& \multicolumn{7}{c}{Engagement (Group I vs Group III)} \\
\cmidrule(r){2-7} 
& \multicolumn{2}{c}{(1) Sex} & \multicolumn{3}{c}{(2) Age} & \multicolumn{2}{c}{(3) Network Degree} \\
\cmidrule(r){2-3} \cmidrule(r){4-5} \cmidrule(r){6-7}
& Female & Male & $<30$ & $30-59$ & $\geq60$ & High \\
\midrule 
$\text{Algorithm}$ $\times$ NR & $0.112^{***}$ & $0.079^{***}$ & $0.084^{***}$ & $0.097^{***}$ &$0.059^{***}$ & $0.057^{***}$ & $0.210^{***}$ \\
& (0.013) & (0.009) & (0.025) & (0.009) & (0.016) & (0.009) & (0.014) \\
\midrule 
$\text{Algorithm}$ $\times$ Diversity & $0.351^{***}$ & $0.253^{***}$ & $0.250^{***}$ & $0.280^{***}$ & $0.257^{***}$ & $0.182^{***}$ & $0.634^{***}$ \\
& (0.043) & (0.031) & (0.071) & (0.029) & (0.070) & (0.030) & (0.042) \\
\midrule 
Observations & 24,332,535 & 40,423,029 & 5,613,756 & 47,240,098 & 11,901,710 & 47,819,777 & 16,935,787 \\
\bottomrule 
\end{tabular}
\footnotetext{Note: Clustered standard errors at the user level in parentheses. Stars indicate the significance level based on $p$ values: ${}^{*} p < 0.1$; ${}^{**} p < 0.05$; ${}^{***} p < 0.01$.}   
\label{tab.hetero_contentdifference}
\end{sidewaystable}
\newpage
\subsection{Heterogeneous Effects of Mechanisms on Users' Tendency to Engage with Novel Information}
We explore users' tendency to engage with novel content under algorithmic versus peer-shared delivery across different user segments, including sex, age, and network degree.
The results presented in Tables~\ref{tab.hetero_socialcue} and \ref{tab.hetero_contentdifference} show that the effect of content differences outweighs that of peer influence across all user segments, particularly for female, younger, and low network-degree users.

\end{appendices}

\end{document}